# On the Existence of Additional (Hydrino) states in the Dirac equation

## ANZOR KHELASHVILI [1,2], TEIMURAZ NADAREISHVILI [1]

[1] Inst. of High Energy Physics, Iv. Javakhishvili Tbilisi State University, University Str. 9, 0109, Tbilisi, Georgia.

[2] St.Andrea the First-called Georgian University of Patriarchy of Georgia, Chavchavadze Ave.53a, 0162, Tbilisi, Georgia.
E-mail : anzor.khelashvili@tsu.ge ;  teimuraz.nadareishvili@tsu.ge

**Abstract**: In case of spinless particles there appear additional (singular) solutions in the framework of relativistic Klein-Gordon equation for Coulomb potential. These solutions obey to all requirements of quantum mechanical general principles. Observation of such states ("hydrino, small hydrogen") should be important for manifestation of various physical phenomena.   In this article the same problem is considered for spin-1/2 particle (electron) in the Dirac equation. It is shown that such kind of solutions really occurs, but the rate of singularity is more higher than in spinless case. By this reason we have no time- independence of total probability (norm). Moreover the orthogonality property is also failed, while the total probability is finite in the certain area of the model-parameters. Therefore, we are inclined to conclude that this additional solution in the Dirac equation must be ignored and restrict ourselves only by normal (standard) solutions.

## 1. Introduction

Time by time there appear articles [1-5] in scientific literature, in which authors consider the so-called additional solutions in the problem of hydrogen-like atoms. These solutions are related to the $1/r^2$ behaving term in potential of the Schrodinger or other equations. They are called by various names: additional, peculiar levels, some times – as a small hydrogen ("hydrino''). It is remarkable to note, that these authors relate these solutions to different physical phenomena, such as a specific  radiation of Galaxy [5], dark matter [5]  or the cold fusion [6,7] in nuclear reactions, etc.

In the earlier papers [ 8,9] we have shown that this kind of solutions really exist in the Klein-Gordon equation for Coulomb potential and their status are established after application of the self-adjoint procedure (SAE). It is interesting that some authors believe that we must have exactly the same situation in the Dirac equation , the so-called Dirac's deep levels (DDL) [10] and  as this belief is nearly overgrowing into the final statement, contrary to them, below we consider this problem more carefully in the framework of the Dirac  radial equation. It turns out that the fulfillment to of all general principles of quantum  mechanics for additional solution is impossible and the final result is pessimistic.

Our approach is based on the radial Dirac equations for Coulomb potential. While the results following from the Dirac equation are often well-known, we will try below to get exhaustive answer.

## 2. Dirac's Radial Equations and Normalization Properties of Radial Wave Functions

As is well-known, to derive radial equations the full Dirac wave function is represented in the following form [11]

$$\Psi(\vec{r}) = \begin{pmatrix} f(r)\Omega_{jlm}(\vec{n}) \\ (-1)^{\frac{1+l-l'}{2}} g(r)\Omega_{jl'm}(\vec{n}) \end{pmatrix}; \quad l = j \pm 1/2, \; l' = 2j - l \quad (1)$$

Here $\Omega_{jlm}$ are spin-spherical harmonics, depended on spherical angles $\vec{n}$. Substituting Eq. (1) into the Dirac equation gives the system of first order radial equations

$$f' + \frac{1+\kappa}{r} f - (E + m - V) g = 0$$
$$g' + \frac{1-\kappa}{r} g + (E + m - V) f = 0 \quad (2)$$

where

$$\kappa = \begin{cases} -(j+1/2) = -(l+1), & \text{if } j = l + 1/2 \\ +(j = 1/2) = l, & \text{if } j = l - 1/2 \end{cases} \quad (3)$$

is the Dirac quantum number, which corresponds to the conserved Dirac matrix

$$K = \beta(\Sigma \cdot l + 1). \quad (4)$$

The behaviour of radial wave function at the origin of coordinates is closely related to the normalization property of wave function. As in the Schrodinger case this behaviour depends on what are required to be finite – probability density, full probability or differential probability [9].

In order not to repeat the well-known procedure [9], below we consider the most general and strong restriction [12] –time-independence of the total probability

$$\frac{d}{dt}\int \psi^+ \psi \, dV = 0 \quad (5)$$

Let us use the time-dependent Dirac equation

$$i\hbar \frac{\partial \psi}{\partial t} = \hat{H}\psi \; ; \; \hat{H} = \vec{\alpha}\vec{p} + \beta m + V(r) \quad (6)$$

Then the condition (5) takes the form

$$-i\int \left[ \psi^+ (H\psi) - (H\psi^+)\psi \right] dV = 0 \quad (7)$$

In other words, time- independence of probability means that the Hamiltonian must be a self- adjoint operator.

Let us begin from the cotinuity equation in the Dirac problem

$$\frac{\partial \rho}{\partial t} + div \boldsymbol{j} = \boldsymbol{0} \tag{8}$$

where
$$\rho = \psi^+\psi, \qquad \boldsymbol{j} = \psi^+\boldsymbol{\alpha}\psi \tag{9}$$

According to the continuity equation, conservation of the norm takes the following form after using the Gauss theorem

$$\frac{\partial}{\partial t}\int_V \psi^+\psi dV = -\int_V div \boldsymbol{j} dV = -\int_S j_N dS \tag{10}$$

where $j_N$ is the normal component of current into the surface.

If we assume that at $r = 0$ the Hamiltonian has a singular point, Gauss' theorem in Eq.(10) is not applicable. We must exclude this point from the integraton volume and surround it by a small sphere of radius $a$. In this case the surface integral is divided into a surface in infinity that encloses the total volume, and the surface of a sphere of radius $a$:

$$\lim_{a\to 0} a^2 \int j_a d\Omega + \int_a j_N dS = 0 \tag{11}$$

In the first integral here we have expressed the surface element of the sphere as $dS = a^2 d\Omega$, where $d\Omega$ is an element of solid angle. Because the wave function must vanish at infinity (bound state). the second term goes to zero. As regards to normal component of $j_N$ (its radial part) using Eq.(1) and taking into account that

$$\alpha_r = \begin{pmatrix} 0 & -i \\ i & 0 \end{pmatrix}, \tag{12}$$

it takes the form

$$j_N = i\left(fg^* - gf^*\right)_{r=a} \tag{13}$$

Because $r$ is confined in a small region one can replace the radial functions by their behaviour near origin $f = \dfrac{u}{r^S}$ and $g = \dfrac{v}{r^S}$, where $u, v$ are regular functions at the origin, we derive

$$\lim_{a\to 0}\frac{a^2}{a^{2S}}\int\left(u\frac{\partial v^*}{\partial r} - v\frac{\partial u^*}{\partial r}\right)_{r=R} d\Omega = 0 \tag{14}$$

This equation is satisfied only if $S < 1$. It follows that radial functions under consideration do not diverge at the origin more quickly than $1/r^S$, with $S < 1$.

## 3. Behaviour for the Coulomb-Like Potentials

For more elucidation let us investigate the behaviour of radial function at the origin for attractive potentials like

$$\lim_{r\to 0} rV(r) = -V_0 = const > 0 \tag{15}$$

In this case, radial equations become (at the origin)

$$f' + \frac{1+\kappa}{r} f - \frac{V_0}{r} g = 0$$
$$g' + \frac{1-\kappa}{r} g + \frac{V_0}{r} f = 0 \qquad (16)$$

As these functions enter here at equal footing, they can be taken in the same form
$$f = Ar^\lambda \quad \text{and} \quad g = Br^\lambda \qquad (17)$$

After substitution into equations we obtain the following condition
$$(\lambda + 1)^2 = \kappa^2 - V_0^2$$

or
$$\lambda = -1 \pm P, \qquad P = \sqrt{\kappa^2 - V_0^2} \qquad (18)$$

It means that the radial wave functions behave at the origin as
$$\lim_{r \to 0} f(r) = a_1 r^{-1+P} + b_1 r^{-1-P} \equiv f_{st} + f_{add}$$
$$\lim_{r \to 0} g(r) = a_2 r^{-1+P} + b_2 r^{-1-P} \equiv g_{st} + g_{add} \qquad (19)$$

If we compare this behaviour to the corresponding result for Schrodinger equation, it appears that now the singularity at origin is grown by factor $r^{-1/2}$. This difference is easy to understand even on the dimensional point of view, because the energy expressions in both cases look like

$$E_{Sch} = \int dV \psi_{Sch}^*(r) \left( \frac{p^2}{2m} + V(r) \right) \psi_{Sch}(r)$$
$$E_{Dir} = \int dV \psi_{Dir}^+(r) (\alpha \cdot p + \beta m + V(r)) \psi_{Dir}(r) \qquad (20)$$

We see that the scale dimensions in these two cases differ exactly by ½ degree.

This fact has important influence on physical picture – the additional solutions behave like $f_{add}(g_{add}) = b_1(b_2) r^{-1-P}$ and the nonrelativistic limit and scale behaviour at the origin do not commute with each others.

Now it is remarkable that this behaviour does not obey to the criterion of time-independence of the norm, because growing at the origin is more rapid than is admissible by this criterion. Because this criterion means the self-adjointess of Hamiltonian, its breaking is not allowed [12].

Nevertheless the total probability is finite, when $0 \leq P < 1/2$
$$\int_a \{f^2 + g^2\} r^2 dr \sim a^{1-2P}; \ a \to 0 \qquad (21)$$

At the same time differential probability in the spherical slice diverges
$$f^2 r^2 dr \sim r^{-2P}, \ r \to 0 \qquad (22)$$

because $P > 0$. It is seen that the answer about existence of additional solution depends on the following point of view: which property is required for the wave function.

In literature the radial equations for functions $rf$ and $rg$ are often considered. In this case instead of system of equations one considered single equation. This is achieved by exclusion of one of the functions from the system (1). For example, if we determine from the second equation of (1) the function $g = \dfrac{1}{E+m-V}\left\{f' + \dfrac{1+\kappa}{r}f\right\}$ and substitute it into the first equation, we are passing to the second order equation for a single function

$$f'' + \frac{2}{r}f' + \frac{V'}{E+m-V}f' + \frac{V'}{E+m-V}\frac{1+\kappa}{r}f +$$
$$+\left[(E-V)^2 - m^2 - \frac{\kappa(\kappa+1)}{r^2}\right]f = 0 \qquad (23)$$

The first two terms of this equation are the radial part of the Laplace operator. One can exclude from these two terms the first derivative, using the known relation [13]:

$$\left(\frac{d^2}{dr^2} + \frac{2}{r}\frac{d}{dr}\right)f = \frac{1}{r}\frac{d^2 F(r)}{dr^2} - 4\pi\delta^{(3)}(\mathbf{r}), \qquad (24)$$

where

$$F(r) = rf(r) \qquad (25)$$

Taking into account that in spherical coordinates $\delta^{(3)}(\mathbf{r}) = \dfrac{\delta(r)}{4\pi r}$, after some transformations we derive the following equation for $F(r)$:

$$\frac{1}{r}\left[\frac{d^2 F(r)}{dr^2} - \frac{\kappa(\kappa+1)}{r^2}\right]F(r) - \delta(r)F(r) +$$
$$+ \frac{V'}{E+m-V}(rF' + \kappa F) + \left[(E-V)^2 - m^2\right]rF(r) = 0 \qquad (26)$$

The term, containing $\delta$-function, has no physical sense and it must be avoided by boundary condition [9]

$$F(0) = 0 \qquad (27)$$

At the same time, in order the product $\delta(r)F(r)$ be a correct defined distribution, one must require that the ordinary function $F(r)$ be a well-defined, infinitely smooth (infinitely differentiable) at $r = 0$. This requirement restricts $F(r)$ to be at $r = 0$ a power like function with integer degree. Therefore, we are faced with the following radial equation

$$\frac{1}{r}\left[\frac{d^2 F(r)}{dr^2} - \frac{\kappa(\kappa+1)}{r^2}\right]F(r) = \frac{V'}{E+m-V}(rF' + \kappa F) + \left[(E-V)^2 - m^2\right]rF(r) = 0 \qquad (28)$$

This equation is also known in scientific literature [14]. But here we want to stress that it acquires its status only after fulfillment of boundary condition (27). Only in this case the solutions of radial equation satisfy the primary total Dirac 3-dimensional equation [9, 15,16]. So the condition (27) is in accordance with the time-independence of norm.

There is also one stronger criterion, which is related to the self-adjointness of Hamiltonian, namely, orthogonality of additional solutions. The solutions derived above, in difference from the Schrodinger or Klein-Gordon equations, do not satisfy orthogonality condition as well [4,8]

$$\lim_{r \to 0}\left(f_k^* g_{k'} - f_{k'} g_k^*\right) = 0; \quad k^2 = E^2 - m^2 \qquad (29)$$

In conclusion, we can say that for the Dirac equation the real existence of "hydrino" is doubtful.